# Pressure-induced phase transition and bandgap collapse in the wide-bandgap semiconductor InTaO$_4$

D. Errandonea[1], C. Popescu[2], A.B. Garg[3], P. Botella[1], D. Martinez-García[1], J. Pellicer-Porres[1], P. Rodríguez-Hernández[4], A. Muñoz[4], V. Cuenca-Gotor[5], and J. A. Sans[5]

[1] Departamento de Física Aplicada-ICMUV, MALTA Consolider Team, Universidad de Valencia, Edificio de Investigación, C/Dr. Moliner 50, Burjassot, 46100 Valencia, Spain

[2] CELLS-ALBA Synchrotron Light Facility, 08290 Cerdanyola, Barcelona, Spain

[3] High Pressure and Synchrotron Radiation Physics Division, Bhabha Atomic Research Centre, Mumbai 400085, India

[4] Departamento de Física, Instituto de Materiales y Nanotecnología, MALTA Consolider Team, Universidad de La Laguna, La Laguna, E-38205 Tenerife, Spain

[5] Instituto de Diseño para la Fabricación y Producción Automatizada, MALTA Consolider Team, Universitat Politècnica de Valencia, 46022 Valencia, Spain

**Abstract:** A pressure-induced phase transition, associated with an increase of the coordination number of In and Ta, is detected beyond 13 GPa in InTaO$_4$ by combining synchrotron x-ray diffraction and Raman measurements in a diamond anvil cell with *ab-initio* calculations. High-pressure optical-absorption measurements were also carried out. The high-pressure phase has a monoclinic structure which shares the same space group with the low-pressure phase (*P2/c*). The structure of the high-pressure phase can be considered as a slight distortion of an orthorhombic structure described by space group *Pcna*. The phase transition occurs together with a unit-cell volume collapse and an electronic bandgap collapse observed by experiments and calculations. Additionally, a band crossing is found to occur in the low-pressure phase near 7 GPa. The pressure dependence of all the Raman-active modes is reported for both phases as well as the pressure dependence of unit-cell parameters and the equations of state. Calculations also provide information on IR-active phonons and bond distances. These findings provide insights into the effects of pressure on the physical properties of InTaO$_4$.

I.  **Introduction**

Ternary oxides of the form $M$TO$_4$ have been extensively studied for their interesting physical properties; in particular, those oxides crystallizing either in the scheelite or the wolframite structure [1 -10]. During the last decade, the most studied members of this family of materials have been the orthotungstates and orthomolybdates, which have many technological applications [11 – 16]. In special, high-pressure studies of $M$TO$_4$ ternary oxides have generated a large amount of attention, being high-pressure an efficient tool to improve the understanding of their physical properties [3 – 12]. Recently, other compounds like perrhenates [17] and orthotantalates [18] have also become the focus of research. One material of particular interest is indium tantalate (InTaO$_4$), a promising candidate as efficient photocatalytic for water splitting, which is a potential source of green energy [19, 20]. At ambient conditions, InTaO$_4$ crystallizes in the wolframite structure (space group: *P2/c*, Z = 2) [21]. This is a monoclinic structure (shown in Fig. 1) in which both In and Ta cations have octahedral oxygen coordination. In the wolframite structure each InO$_6$ (TaO$_6$) octahedron shares two edges with neighboring InO$_6$ (TaO$_6$) octahedral units forming a zig-zag chain and it is connected by the six corners with TaO$_6$ (InO$_6$) neighbors [22, 23]. InTaO$_4$ has been proposed to be a wide-bandgap semiconductor

In spite of the extensive interest in InTaO$_4$, the understanding of its fundamental physical properties is scarce yet. Basically, only the crystal structure of this oxide is known. Raman- and infrared (IR)-active phonons have not been studied and there are important discrepancies regarding the electronic bandgap [19, 24, 25, 26], for which a value ranging from 2.6 to 4.1 eV is reported. In contrary with other ternary oxides, to our knowledge, no high-pressure (HP) studies are available in the literature for InTaO$_4$.

Here we will report a combined experimental and theoretical study of InTaO$_4$ at ambient and high-pressure. X-ray diffraction (XRD), Raman spectroscopy, and optical absorption experiments have been carried out up to 22 GPa. First-principle calculations have also been performed. We will report an accurate determination of phonons and the electronic bandgap for the wolframite phase of InTaO$_4$, supported by the agreement between theory and experiments. We also report evidence of the existence of a first-order phase transition beyond 13 GPa to another monoclinic structure that can be described with space group *P2/c*. The transition involves a collapse of the volume and an atomic rearrangement, which has important consequences on the physical properties of InTaO$_4$; e.g. the bandgap energy (E$_g$) collapses 1.3 eV at the transition. The evolution under pressure of unit-cell parameters, Raman and IR modes, and E$_g$ is also reported for the low- and high-pressure phases. The equation of state (EOS) is also determined. The reported studies have enabled us to improve the understanding of the physical properties of InTaO$_4$ and other ternary oxides and their behavior under compression.

## II. Experimental details

Polycrystalline InTaO$_4$ was synthesized by a ceramic route starting from pre-dried In$_2$O$_3$ and Ta$_2$O$_5$ (purity > 99.9%). The purity of the synthesized material was confirmed by Energy-dispersive x-rays spectroscopy carried out in transmission-electron microscope (TEM) operated at 200 KeV at the SC-SIE from Universitat de Valencia. It was also verified by powder XRD measurements, using Cu K$_\alpha$ radiation, that the samples were single-phased and presented the wolframite-type structure (*P2/c*). The unit-cell parameters determined to be *a* = 4.826(4) Å, *b* = 5.775(5) Å, *c* = 5.155(5) Å, and β = 91.37(3)º, in very good agreement with the reported values in the literature [21, 24].

High-pressure measurements were performed using a membrane diamond-anvil cell and a 16:3:1 methanol-ethanol-water mixture as pressure-transmitting medium. Pressure was measured using the ruby scale [27]. In the XRD experiments, in addition to the sample, we loaded close to it a grain on Cu in order to use it as a second pressure scale [28]. Special caution was taken during DAC preparation to avoid sample bridging between diamonds [29].

HP XRD experiments were performed at the MSPD-BL04 beamline of ALBA synchrotron [30], using a monochromatic beam of wavelength 0.4246 Å which was focused to a 15 µm × 15 µm spot (full-width half maximum) using Kirkpatrick-Baez mirrors. At each pressure we collect XRD patterns from both the sample (used to determine the crystal structure) and the sample plus copper (used to determine pressure). Diffraction images were collected using a Rayonix CCD detector with an exposure time of 10-30 seconds. The FIT2D software [31] was used to calibrate sample to detector distance, detector tilt and to integrate the two-dimensional diffraction images to standard one dimensional intensity versus 2θ plot. The structural analysis was performed with GSAS software package [32].

Ambient and HP Raman spectra were collected in the backscattering geometry using a 514.5 nm Ar$^+$ laser, a Jobin−Yvon TRH1000 spectrometer, and a thermoelectric-cooled multichannel CCD detector [33]. The set-up was calibrated using plasma lines of the argon ion laser. The spectral resolution of the system was below 2 cm$^{-1}$. A laser power of less than 10 mW before the DAC was used to avoid sample heating. Two experimental runs were carried out with similar results.

For optical absorption studies, we used 10-µm-thick polycrystalline platelets, which were obtained by compressing the InTaO$_4$ powder to 1 GPa using a hydraulic press equipped with Bridgman anvils [34]. Measurements in the visible−near-infrared range were made with an optical setup that consisted of a tungsten lamp, fused silica lenses, reflecting optics objectives, and a visible−near-infrared spectrometer [35]. The optical absorption spectra were obtained from the transmittance spectra of the sample, which were recorded using the sample-in, sample-out method [36, 37].

**III.    Ab Initio simulations**

*Ab initio* total-energy calculations, based on density functional theory (DFT) [38] have been performed. The Vienna *ab initio* simulation package (VASP) [39] has been used with the pseudopotential method and the projector augmented wave scheme (PAW) [40] in order to include the full nodal character of the all electron charge density in the core region. Due to the presence of the oxygen atoms in the structure, the basis set of plane waves was extended up to a large energy cutoff of 520 eV, to achieve an accurate description of the electronic properties. The exchange-correlation energy was considered in the generalized gradient approximation (GGA) with the PBEsol prescription [41]. A dense special k-points sampling for the Brillouin Zone (BZ) integration was used to have very well converged energies and forces. At selected volumes, the structures considered in this study were fully relaxed to their optimized configuration through the calculation of the forces and the stress tensor. In the optimized configurations, the forces on the atoms were less than 0.006 eV/Å and the deviation of the stress tensor from a diagonal hydrostatic form was smaller than 1 kbar (0.1 GPa). The resulting set of energy, volume, and energy data (E, V, P) was fitted with a fourth-order Birch-Murnaghan equation of state [42] to evaluate the equilibrium volume ($V_0$), the bulk modulus ($B_0$), and its pressure derivatives ($B_0'$, $B_0''$). The simulations were performed at zero temperature (T=0), then the stable structures and the transition pressure can be determined analyzing the enthalpy as function of pressure, H(P).

Lattice-dynamics calculations were performed at the zone center (Γ point) of the BZ to study and further analyze the vibrational modes under pressure. The construction of the dynamical matrix at the Γ point of the BZ, employing the direct method, involves separate calculation of the forces in which a fixed displacement from the equilibrium configuration of the atoms within the primitive cell is considered. Due to the crystal symmetry the number of independent displacements in the analyzed structure is reduced. The frequencies of the normal modes were obtained from the diagonalization of the dynamical matrix. The calculations allow also identifying the symmetry and eigenvectors of the vibrational modes at the Γ point. Electronic band structure calculations, at several pressures, were performed within the first Brillouin zone along the high-symmetry direction Γ-B-D-Z-Γ-Y-C-Z, for the low- and high-pressure structures [3].

## IV. Results and discussion

### A. HP x-ray diffraction

Fig. 2 shows a selection of powder XRD patterns recorded at different pressures. On compression we found that up to 12.8 GPa all the diffraction patterns can be assigned to the low-pressure wolframite structure. Only a shift toward large angles is observed in the Bragg peaks due to the unit-cell contraction caused by the increase of pressure. Fig. 3 shows in the top panel a diffraction pattern measured at 1.4 GPa together with a Rietveld refinement assuming the wolframite structure (space group *P2/c)*. In the refinement, not only the unit-cell parameters, but also the atomic positions [In: (2f: 1/2, y,1/4), Ta: (2e: 0, y, 1/4), $O_1$: (4g: x, y, z), and $O_2$: (4g: x, y, z)] were considered as free parameters. Other parameters as occupancy and isotropic displacement parameters were constrained as usually done when analyzing HP XRD data [43]. Various R-factors of the refinements and structural parameters obtained at 1.4

GPa are given in Table I. Similar quality of Rietveld refinements were obtained up to 12.8 GPa. In Table I it can be seen that the refined atomic positions at 1.4 GPa are very similar to those obtained at ambient pressure [44]. In addition, we found that up to 12.8 GPa the effect of pressure on the atomic positions is comparable with the uncertainty of the experiments.

The XRD pattern recorded at 13.1 GPa shows several additional weak reflections which become more prominent at higher pressure (see Fig. 2). In addition, the Bragg peaks of the low-pressure phase disappear beyond 15.9 GPa. This fact suggests that wolframite-type $InTaO_4$ undergoes structural changes at 13.1 GPa, coexisting the low- and high-pressure phases from 13.1 to 15.9 GPa. All the reflections observed in the XRD patterns recorded beyond 15.9 GPa could be linked to a single phase. In particular, we found that all the Bragg reflections from the high-pressure phase could be accounted also by space group *P2/c*. Subsequently, we determined the crystal structure of the HP phase by the Rietveld method. In the refinements, we used as starting values the unit-cell parameters obtained using the Le Bail method. For the crystalline structure we employed a model based on the positional coordinates obtained from our *ab initio* calculations. Then, a similar procedure than for the Rietveld refinements of the low-pressure phase was used, but in this case, due to the lower quality of XRD patterns, the oxygen positions were not refined. Fig. 3 shows in the bottom panel the results of the refinement made from the XRD pattern measured at 20.4 GPa. Details on the refined structure from this XRD pattern can be found in Table I. There it can be seen that the monoclinic $\beta$ angle is very close to 90º, so it could be said that the HP phase is "pseudo-orthorhombic". The obtained R-values (see Table I) and small residuals (see Fig. 3) suggest that the proposed monoclinic structure is a good candidate for the HP phase of $InTaO_4$. Therefore, both the low- and high-pressure phases can be described by the same space group, with the atoms occupying the same Wyckoff positions, i.e. the transition does not involve a change in the symmetry of $InTaO_4$. However, the transition involves a volume collapse of approximately 10% and an increase in the coordination of both In and Ta atoms from 6 to 8. These facts indicate that the transition is a first-order transformation. In the HP phase, at 20.4 GPa, each coordination polyhedron has four bond distances with multiplicity of two; In-O = 2.038(9) Å, 2.151(9) Å, 2.225(9) Å, and 2.366(9) Å; and Ta-O = 1.855(9) Å, 1.966(9) Å, 2.267(9) Å, and 2.412(9) Å. As can be seen in Fig. 1, the HP phase is much more compact than the low-pressure phase, consisting of a network of edge-connect $InO_8$ and $TaO_8$ dodecahedra. All the XRD patterns measured from 16.5 GPa to the highest pressure reached in the experiments (22.5 GPa) can be identified with the monoclinic HP phase described above. On decompression the phase transition is reversible. We recovered a pure wolframite phase as shown in Fig. 2 with the pattern measured at 1 GPa upon pressure release. In particular, the HP phase was observed on decompression up to 13.1 GPa and the low-pressure phase comes back at 10.8 GPa, so the hysteresis of transition is about 2.3 GPa.

The pressure dependence of unit-cell parameters of the two phases of $InTaO_4$ have been extracted from the Rietveld refinements and are plotted in Fig. 4. The compression of the low-pressure wolframite phase is anisotropic as observed in other wolframite-structured oxides [22, 45]. Particularly, in $InTaO_4$, the *b*-axis is the most compressible axis. The reason for being less stiff compared to the other axes is the presence of an inter-polyhedral empty space along the (010) direction. Other fact to remark on the effect of pressure in the low-pressure phase is that the β angle follows a non-linear behavior. The β angle gradually decreases under compression up to 3 GPa and then starts increasing up to the transition pressure. The origin of this behavior is related to the fact that up to 3 GPa the compression of InTaO$_4$ is mainly dominated by the compression of the InO$_6$ and TaO$_6$ octahedra. However, beyond 3 GPa part of the volume contraction is accommodated by a tilting of the octahedral units, which favor the increase of the β angle under compression.

Results on the compressibility of the high-pressure phase can also be seen in Fig. 4. In this case, we found that the compressibility of the three axes is similar and much smaller than in the low-pressure phase. On the other hand, the β angle increases slightly under compression from 90.3º at 14.2 GPa to 90.5º at 22.5 GPa. Fig. 4 also shows the pressure-volume data obtained for both phases. These results have been analyzed using a third-order Birch–Murnaghan (BM) equation of state (EOS) [46] and the EosFit7software [47]. From the EOS fit, we obtained the ambient pressure bulk modulus (B$_0$) and its pressure derivative B$_0$' as well as the unit-cell volume at ambient pressure (V$_0$). The EOS parameters are given in Table II together with the implied values for the second pressure derivative of the bulk modulus, B$_0$″ [48]. These results suggest that wolframite-type InTaO$_4$ is less compressible than wolframite-type orthotungstates [22]. In addition, the HP phase of InTaO$_4$ is less compressible than its low-pressure phase, which is consistent with the increase of density associated to the phase transition.

**B. Raman experiments**

Fig. 5 shows selected Raman spectra of InTaO$_4$ at different pressures up to 19 GPa. The point group of the low-pressure wolframite structure ($C_{2h}$) gives rise to 8 A$_g$ modes and 10 B$_g$ modes that are Raman active. In our experiments, we were able to follow the 18 Raman-active modes of the wolframite phase under pressure. The wavenumbers of the eighteen modes are given in Table III. Mode assignment has been done according to the calculations reported below. The Raman spectra of InTaO$_4$ shows the typical Raman spectrum of wolframites with four high-frequency modes, two $A_g$ and two $B_g$, which are separated by a phonon gap from the rest of the Raman modes. In wolframites (e.g. $CdWO_4$ [5] and $MnWO_4$ [49]), it is also generally possible to observe the presence of two Raman modes around 500 cm$^{-1}$, one $A_g$ and one $B_g$, with the rest of the twelve modes below 421 cm$^{-1}$. This feature is also observed in the Raman spectrum of the low-pressure phase of $InTaO_4$. Fig. 6 displays the pressure dependence of all Raman modes. The evolution of the Raman modes with pressure can be considered nearly linear for all of them. The pressure coefficients for each mode (dω/dP) are also given in Table III together with the Grüneisen parameters ($\gamma = \frac{B_0}{\omega_0}\frac{\partial \omega}{\partial P}$). The mode with the highest γ is the $B_g$ mode observed at 421 cm$^{-1}$. Exactly the same behavior has been previously found by the equivalent mode in wolframite-type $MgWO_4$ [3] and $MnWO_4$ [49]. Therefore, apparently, this $B_g$ mode is most sensitive Raman mode of wolframites to pressure; i.e. the Raman vibration to which more energy is transferred during compression. In addition, up to 13.1 GPa we can observe a frequency shift to higher frequencies with compression of most of the modes, but with two modes showing a gradual softening (see Table III). Some of the modes merge under compression due to their different pressure coefficients as can be seen in Figs. 5 and 6.

At 13.4 GPa new Raman-active modes appear and coexist with those of the wolframite phase up to 17.3 GPa. The appearance of these new peaks at 13.4 GPa (denoted by asterisks in Fig. 5) is interpreted as a confirmation of the onset of the structural phase transition detected in the XRD experiments. The coexistence range is consistent with the results obtained from XRD experiments. Again the transition is gradual as weak peaks of the low-pressure phase are still detected at 17.3 GPa (denoted by $ in the figure). The high-pressure phase is stable up to 20.5 GPa and the transition is again found to be reversible, which is illustrated in Fig. 5 by a Raman spectrum

measured at 2 GPa under pressure release. The obtained pressure dependence of the Raman modes of the HP phase is shown in Fig. 6. The frequencies and pressure coefficients are summarized in Table IV. Since the HP phase has the same point group and identical number of atoms per formula unit than wolframite, again 8 $A_g$ and 10 $B_g$ modes are expected. We observed all these modes. The mode assignment of Table IV was done following our calculations. In the case of the HP phase, again there are twelve modes located at low frequencies. In this case, they are below 500 cm$^{-1}$. However, the mode distribution is slightly different than in the wolframite phase. In particular, in the HP phase there are two modes above 800 cm$^{-1}$ and four modes from 560 to 700 cm$^{-1}$. The redistribution of the high-frequency modes is consistent with the coordination change determined from XRD experiments [50]. Regarding the pressure coefficients, we found that in the HP phase all of them are positive (i.e. all the modes harden). We also found that in the HP phase the low-frequency modes have larger Grüneisen parameters than in the low-pressure phase. As in the low-pressure phase, in the HP phase again the mode with the highest Grüneisen parameter is an intermediate frequency $B_g$ mode; the mode with wavenumber at 454 cm$^{-1}$ (see Table IV).

### C. Optical-absorption experiments

The absorption coefficient (α) of the two phases of InTaO$_4$ at different pressures is shown in Figs. 7 and 8. For the low-pressure wolframite phase, at ambient pressure, the absorption spectrum shows a steep absorption, which corresponds to the overlapping of the fundamental absorption plus a low-energy absorption band. This low-energy absorption corresponds to the typical Urbach tail observed in related ternary oxides [51] plus a small diffusion of light between the grains of the polycrystalline sample used in the measurements. The nature of the low-energy tail has been the subject of considerable debate [52] and is beyond the scope of this work. In order to determine the nature and energy of the fundamental band gap, we analyzed the absorption spectra using a Tauc plot [53]. We found that for high energies $\sqrt{\alpha E}$ is proportional to the photon energy (E), as can be seen in Fig. 7 (bottom), which supports that InTaO$_4$ is an indirect-band-gap semiconductor. The value determined for the bandgap energy is E$_g$ = 3.79(5) eV. This value is slightly smaller but consistent with the value reported by Malingowski *et al.* (Eg = 3.96 eV) [19], confirming that InTaO$_4$ is a wide-gap semiconductor. Previous [24] and present first-principle calculations also support the conclusion extracted from the experiments. Therefore, the small bandgap (Eg = 2.6 eV) previously reported [25, 26] was probably an underestimated value. This could be caused by a wrong interpretation of the low-energy Urbach tail, which in the diffused reflectance measurements reported earlier, was assumed to be the fundamental intrinsic absorption [25, 26].

Under compression, we found that in the wolframite phase of InTaO$_4$ there is a blue shift of absorption spectrum with compression (see Fig. 7). From the spectra measured from ambient pressure to 12.9 GPa we obtained the pressure dependence of E$_g$, which is shown in Fig. 7. Up to 5 GPa E$_g$ moves at a rate of 20 meV/GPa. Above this pressure, there is a decrease of the pressure coefficient which becomes dE$_g$/dP = 15 meV/GPa. This feature can be explained by a band crossing, as it will be discussed in the analysis of the band structure calculations. When increasing the pressure from 12.9 to 13.5 GPa we observed that InTaO$_4$ changes from colorless to yellow. This color change is caused by a bandgap collapse, which can be clearly seen by comparing Fig. 7 with Fig. 8. The occurrence of the abrupt change in the absorption spectrum is consistent with the structural phase transition found by Raman and XRD experiments. For the HP phase we found that $(\alpha E)^2$ is proportional to the photon energy (see Fig. 8); this indicates that the HP phase behaves as a direct bandgap material. From the measurement made up to 22.5 GPa, we obtained the pressure dependence for $E_g$ in the HP phase. In this phase, the bandgap redshift under compression with $dE_g/dP = -8$ meV/GPa. The collapse of $E_g$ at the phase transition is estimated to be 1.3 eV.

### D. Ab Initio Calculations

Fig. 9 shows the energy−volume and enthalpy-pressure curves obtained from our calculations for the wolframite phase and the HP phase. We found that wolframite is the stable phase of $InTaO_4$ at ambient conditions. The calculated structure is given in Table V. The agreement with the experiments is excellent. From our calculation we obtained the pressure evolution of the unit-cell parameters for the wolframite phase. The results are compared in Fig. 4 with the experiments. The agreement is quite good. Calculations also found that the β angle follows a non-linear behavior, confirming the experimental results.

Our simulations also provide information of the effect of pressure in the atomic positions. Atomic positions at 13.4 GPa are given in Table V for comparison with the ambient pressure values. We conclude that not only Ta and In atoms moves under compression towards a higher symmetry position, but also the oxygen atoms move considerably. From the calculations we obtained information on the contraction of the $InO_6$ and $TaO_6$ octahedra. The last polyhedron has a volume of 10.135 Å$^3$ at ambient pressure and contracts to 9.410 Å$^3$ at 13.4 GPa; i.e. the volume of the polyhedron is reduced by 7%. In addition, it gets more distorted increasing the distortion index (defined by Bauer [54]) from 0.061 to 0.072. In the case of $InO_6$, the volume of the octahedron is reduced from 13.692 Å$^3$ to 13.093 Å$^3$. Consequently, it undergoes a 5% contraction. In this case the octahedron becomes more symmetric, decreasing the distortion index from 0.031 to 0.010.

Regarding the HP phase, our calculations found that, after the relaxation and optimization of the geometry, the monoclinic structure observed in the experiment becomes pseudo-orthorhombic with the β angle basically identical to 90º. The structure can be described by a translationgleiche supergroup of *P2/c*, the orthorhombic space group *Pcna*. Therefore, we consider that the monoclinic structure found in the experiments is a slight distortion of the theoretically predicted orthorhombic structure. Indeed the orthorhombic structure explains all the Bragg peaks present in the XRD pattern but not the weak peak observed at low angles near 2θ = 4.5º. This peak is explained only by reducing the symmetry from orthorhombic to monoclinic. The distortion of the structure needed to account all XRD peaks is very small (see Tables I and V). Indeed, according to our calculations, only a small deviatoric stress of 0.4 GPa is needed to make the experimental HP monoclinic structure as stable as the theoretical orthorhombic structure. The simulated HP monoclinic structure is more stable than wolframite structure beyond 20 GPa. The small difference between the experimental and theoretical transition pressures might be a consequence of the fact that calculations were carried out at zero temperature and experiments at room temperature (and the fact that calculations do not include kinetics effects). Details on the calculated HP structures (orthorhombic and monoclinic) are given in Table V. The calculated pressure dependence of the unit-cell parameters are shown in Fig. 4, being in good agreement with the experiments. The only difference is seen on the β angle which remains close to 90º at all pressure in the calculations, but increase slightly in experiments. We think this is caused by the slight increase of deviatoric stresses as pressure increases.

To conclude the discussion of the structural calculations we would like to mention that we fit the calculated pressure dependence of the unit-cell volume using a third-order BM EOS. The results are given in Table II. We found there is a tendency of the calculations to give bulk modulus 10% lower than the one determined from the experiments. The opposite behavior is observed for the unit-cell volume, which is slightly higher in the calculations. These facts are well known and typical of GGA functionals.

In addition to the structural calculations we have also performed lattice-dynamical calculations for the two phases of $InTaO_4$. Tables III and IV compare the calculated frequencies and pressure coefficients for the Raman modes with the experimental ones. The agreement between calculations and experiments for the low-pressure phase is very good and better than obtained in wolframite-type orthotungstates [5, 55, 56]. Calculations have provided the mode assignment used to identify the symmetry ($A_g$ or $B_g$) of the eighteen Raman modes. In wolframite-type tungstates, due to the fact that the $WO_6$ octahedron is a rigid and uncompressible unit, it is common to describe the Raman modes as internal and external modes with respect to this octahedron. We tried to apply the same distinction to wolframite-type $InTaO_4$ by visualizing the atomic displacements associated to each Raman mode using J-ICE [57]. However, we found that the fact that the size and compressibility of $InO_6$ and $TaO_6$ octahedral units are similar prevents the use of this simple distinction between modes. Indeed, only the highest frequency mode can be identified as a pure internal mode of the $TaO_6$ octahedron, corresponding to a symmetric-stretching vibration. Note that the frequency of this mode is at least 50 cm$^{-1}$ smaller than the frequency of the equivalent mode of tungstates even if Ta is slightly lighter than W. Usually, such a frequency decrease is associated to a decrease in bond strength, which is consistent with the fact that the Ta-O bonds are more than 10% larger than the W-O bonds.

From our calculations we also obtained the frequencies of the infrared-active modes of the wolframite structure (7 $A_u$ + 10 $B_u$, there is also an $A_u$ silent mode) whose calculated frequencies and pressure coefficients are reported in Table VI. Unfortunately, there are no experimental data available to compare with. We hope our calculations will trigger the curiosity for IR spectroscopy measurements in InTaO$_4$. According to our calculations, the IR modes have a similar frequency distribution than the Raman modes; however, the highest frequency mode is only at 763 cm$^{-1}$. In addition, the IR modes show a similar sensitivity to pressure than the Raman modes. It is important to notice that there are five IR modes with negative pressure coefficients (see Table VI), some of them with a large pressure coefficient. The presence of modes that gradually soften under compression usually is observed in oxides that undergo pressure-driven transitions at low pressure [58], as observed in the case of InTaO$_4$.

Regarding the IR phonons of the HP phase, they are also summarized in Table VI. They cover a wider frequency range than the IR modes of the wolframite phase, but they have a qualitative similar frequency distribution. In particular, the highest frequency mode has A$_u$ symmetry and corresponds to stretching vibrations of the Ta-O bonds. Table VI also provides the pressure coefficients. In the HP phase there is only one IR mode which gradually softens under compression, this is an A$_u$ mode with wavenumber 197.8 cm$^{-1}$.

We will discuss now the band structure of InTaO$_4$. The calculated band structure at different pressure is shown in Fig. 10. At ambient pressure, the top of the valence band is at the Y point of the Brillouin zone (BZ) [3]. However, there is a second maximum very close in energy at the Z point of the BZ. On the other hand, the minimum of the conduction band is in a point in the Γ-B direction of the BZ. Consequently, wolframite-type InTaO$_4$ is an indirect gap material, as suggested by the experiments. Calculations predict at ambient pressure E$_g$ to be 3.74 eV, which is in excellent agreement with our measurements. As pressure increases we observe two facts. First, the conduction band moves towards higher energy. Second, the top of the valence band is less sensitive to pressure. However, pressure drives a band-crossing, becoming around 6 GPa the maximum at the Z point of the BZ the one with the highest energy. The different pressure dependence of the maxima at Y and Z not only causes the band-crossing already described, but also the change on the pressure dependence of $E_g$ already described in the experiments. The calculated $E_g$ versus pressure is shown in Fig. 7. The agreement is excellent, providing calculations a simple explanation to the experimental behavior of $E_g$.

Regarding the HP phase, our calculations indicate that the HP phase of $InTaO_4$ is a direct gap semiconductor, with both the maximum of the valence band and the minimum of the conduction band at the $\Gamma$ point of the BZ. The direct bandgap nature of the HP phase agrees with the conclusion extracted from the experiments. The calculated value of $E_g$ at 17.5 GPa is 2.25 eV (for the simulated HP monoclinic structure). This value is 0.4 eV smaller than the measured value, but the difference is within the range of the typical underestimation of $E_g$ by DFT calculations. On the other hand, calculations provide a similar evolution of $E_g$ with pressure as from experiments. In particular, the small red-shift of the bandgap is caused by the fact that the top of the valence band moves faster towards higher energies than the bottom of the conduction band.

To better understand the pressure evolution of $E_g$ in the low- and high-pressure phases, we analyze the total and partial electronic density-of-states of $InTaO_4$ at different pressures. Fig. 11 compares the total density-of-states of the low-pressure phase at ambient pressure and 7.3 GPa and of the HP phase at 20.5 GPa. In the figure, it can be seen that, in low-pressure structure, the O $2p$ states dominate the upper part of the valence bands. On the other hand the Ta $5d$ states and In $5s$ dominate the lower conduction bands of the low-pressure phase. Under compression, the top of the valence band moves gradually towards lower energies leading to the blue-shift experimentally observed for $E_g$. Regarding the HP phase, the orbital contribution to the valence band resembles very much that of the low-pressure wolframite structure. However, there are qualitative differences in the bottom of the conduction band, which is only dominated by Ta $5d$ states. In this case, the shift with pressure of the bottom of the conduction band and top of the valence band are similar, moving the valence band under compression slightly faster towards high energies than the conduction band, which cause the observed small red-shift of $E_g$.

**V.     Conclusions**

We have performed high-pressure XRD, Raman, and optical-absorption measurements as well as *ab initio* calculations on $InTaO_4$. Changes in the structural, lattice-dynamics, and optical properties between 13 and 17 GPa indicate the occurrence of a pressure-driven phase transition, which involves a significant change in the electronic structure of this material. Density functional calculations confirm the experimental findings and help to understand them. The observed phase transition is reversible and involves a coordination increase for In and Ta. The HP phase has the same monoclinic symmetry as the low-pressure wolframite phase. The evolution of unit-cell parameters, Raman modes, and bandgap energy is reported for the two phases of $InTaO_4$. Calculations also provide information on IR-active phonons and bond distances. The reported results contribute to improve the understandings of the effects of pressure in the physical properties of ternary oxides, which exhibits a number of interesting phase transitions as a function of pressure. Metallic and even superconductivity behavior has been predicted for some of them [17]. Our findings provide insights into the effects of pressure on the physical properties of $InTaO_4$, and they may also help to understand novel properties of other ternary oxides. The study of the electronic structure and the effects of pressure on it might have implications for improving of the water splitting activity of InTaO$_4$.

**Acknowledgments**

This work was partially supported by the Spanish MINECO under Grants MAT2013-46649-C04-01/02/03 and MAT2015-71070-REDC (MALTA Consolider). XRD experiments were performed at MSPD-BL04 beamline at ALBA Synchrotron with the collaboration of ALBA staff. We thank the technical support in TEM measurements provided by S. Agouram from SC-SIE at Universitat de Valencia.

**Table I:** Wolframite structure of InTaO$_4$ at 1.4 GPa (top) and HP structure of InTaO$_4$ at 20.4 GPa (bottom). Both structures belong to space group *P2/c*. For the low-pressure (HP) phase the goodness of fit parameters are: R$_{wp}$ = 6.9 % and R$_P$ = 4.6 % (R$_{wp}$ = 8.8 % and R$_P$ = 6.6 %). The Wyckoff position for each atom is indicated in the first column. For the HP phase only the In and Ta position have been refined.

| \multicolumn{4}{c}{$a$ = 4.818(4) Å, $b$ = 5.760(5) Å, $c$ = 5.146(5) Å, $\beta$ = 91.35(3)°} | | | |
|---|---|---|---|
| Atom | x | y | z |
| In (2f) | 0.5 | 0.6780(16) | 0.25 |
| Ta (2e) | 0 | 0.1738(4) | 0.25 |
| O$_1$ (4g) | 0.2295(19) | 0.8999(18) | 0.4399(25) |
| O$_2$ (4g) | 0.2440(21) | 0.6269(19) | 0.8939(23) |
| \multicolumn{4}{c}{$a$ = 4.872(4) Å, $b$ = 5.082(5) Å, $c$ = 4.909(5) Å, $\beta$ = 90.43(3)°} | | | |
| Atom | x | y | z |
| In (2f) | 0.5 | 0.7713(21) | 0.25 |
| Ta (2e) | 0 | 0.2286(18) | 0.25 |
| O$_1$ (4g) | 0.23856 | 0.94519 | 0.51851 |
| O$_2$ (4g) | 0.23900 | 0.44598 | 0.47955 |

**Table II:** EOS parameters for different structures determined from present experiments and calculations.

| Phase | | V$_0$ (Å$^3$) | B$_0$ (GPa) | B$_0$' | B$_0$''(GPa$^{-1}$) |
|---|---|---|---|---|---|
| LP | Exp. | 143.9(7) | 179(5) | 6.2(7) | -0.076(8) |
| LP | Theo. | 145.6 | 170.8 | 3.83 | -0.0219 |
| HP | Exp. | 131.4(9) | 202(9) | 6.2(9) | -0.054(8) |
| HP | Theo. | 134.9 | 181.1 | 4.01 | -0.0215 |

**Table III:** Raman modes at ambient pressure, pressure coefficients, and Grüneisen parameters for the low-pressure phase of InTaO$_4$.

| Mode | Experiment | | | Theory | |
|---|---|---|---|---|---|
| | ω (cm$^{-1}$) | dω/dP (cm$^{-1}$/GPa) | γ | ω (cm$^{-1}$) | dω/dP (cm$^{-1}$/GPa) |
| B$_g$ | 110 | 0.59 | 0.96 | 106 | 0.64 |
| A$_g$ | 115 | 0.54 | 0.84 | 110 | 0.24 |
| B$_g$ | 144 | 0.90 | 1.12 | 137 | 0.85 |
| B$_g$ | 165 | 0.23 | 0.25 | 163 | 0.16 |
| B$_g$ | 186 | -0.15 | -0.14 | 177 | -0.68 |
| A$_g$ | 219 | 1.47 | 1.20 | 214 | 1.92 |
| A$_g$ | 276 | -0.33 | -0.21 | 263 | -0.57 |
| B$_g$ | 286 | 2.10 | 1.31 | 278 | 1.30 |
| B$_g$ | 301 | 1.56 | 0.93 | 287 | 1.38 |
| A$_g$ | 365 | 0.76 | 0.37 | 349 | 1.19 |
| A$_g$ | 413 | 2.20 | 0.95 | 394 | 2.00 |
| B$_g$ | 421 | 4.19 | 1.78 | 407 | 4.27 |
| B$_g$ | 488 | 3.78 | 1.39 | 476 | 3.78 |
| A$_g$ | 520 | 3.14 | 1.08 | 510 | 3.40 |
| B$_g$ | 651 | 5.08 | 1.40 | 635 | 5.55 |
| A$_g$ | 661 | 4.71 | 1.28 | 650 | 5.07 |
| B$_g$ | 683 | 5.18 | 1.36 | 677 | 5.59 |
| A$_g$ | 829 | 4.12 | 0.89 | 804 | 4.95 |

**Table IV:** Raman modes, pressure coefficients, and Grüneisen parameters for the high-pressure phase of $InTaO_4$. Experiments correspond to 19 GPa and theory to 20.5 GPa.

| Mode | Experiment | | | Theory | |
| --- | --- | --- | --- | --- | --- |
| | $\omega$ (cm$^{-1}$) | d$\omega$/dP (cm$^{-1}$/GPa) | $\gamma$ | $\omega$ (cm$^{-1}$) | d$\omega$/dP (cm$^{-1}$/GPa) |
| $A_g$ | 94 | 0.97 | 2.08 | 81.7 | 0.26 |
| $B_g$ | 135 | 3.15 | 4.71 | 137.6 | 1.12 |
| $B_g$ | 166 | 2.73 | 3.32 | 138.8 | 0.47 |
| $B_g$ | 190 | 1.52 | 1.62 | 175.1 | 0.52 |
| $B_g$ | 223 | 2.46 | 2.22 | 178.8 | 2.65 |
| $A_g$ | 285 | 5.25 | 3.72 | 182.7 | 2.72 |
| $A_g$ | 303 | 4.04 | 2.69 | 221.7 | 1.84 |
| $A_g$ | 319 | 1.18 | 0.74 | 264.2 | 1.80 |
| $B_g$ | 376 | 2.11 | 1.13 | 390.5 | 1.53 |
| $A_g$ | 403 | 0.76 | 0.38 | 404.3 | 1.19 |
| $B_g$ | 454 | 7.18 | 3.19 | 446.4 | 8.78 |
| $B_g$ | 495 | 3.34 | 1.36 | 451.4 | 2.78 |
| $A_g$ | 555 | 3.07 | 1.12 | 560.9 | 2.69 |
| $B_g$ | 600 | 6.62 | 2.23 | 586.1 | 2.65 |
| $B_g$ | 634 | 4.24 | 1.35 | 710.4 | 3.17 |
| $A_g$ | 682 | 5.67 | 1.68 | 731.7 | 2.38 |
| $B_g$ | 830 | 4.91 | 1.19 | 763.5 | 2.22 |
| $A_g$ | 875 | 6.90 | 1.59 | 782.7 | 1.75 |

**Table V:** Calculated wolframite structure of InTaO$_4$ at ambient pressure (top), 13.4 GPa (second from top), and HP orthorhombic (third from top) and monoclinic (bottom) structures of InTaO$_4$ at 20.5 GPa.

| \multicolumn{4}{c}{$a = 4.842$ Å, $b = 5.808$ Å, $c = 5.170$ Å, $\beta = 91.23°$} | | | |
|---|---|---|---|
| Atom | x | y | z |
| In (2f) | 0.5 | 0.67197 | 0.25 |
| Ta (2e) | 0 | 0.17638 | 0.25 |
| O$_1$ (4g) | 0.22479 | 0.89762 | 0.43565 |
| O$_2$ (4g) | 0.24012 | 0.62001 | 0.89939 |
| \multicolumn{4}{c}{$a = 4.775$ Å, $b = 5.639$ Å, $c = 5.050$ Å, $\beta = 91.15°$} | | | |
| Atom | x | y | z |
| In (2f) | 0.5 | 0.68064 | 0.25 |
| Ta (2e) | 0 | 0.18526 | 0.25 |
| O$_1$ (4g) | 0.22163 | 0.90613 | 0.45878 |
| O$_2$ (4g) | 0.24308 | 0.60738 | 0.89356 |
| \multicolumn{4}{c}{$a = 4.90755$ Å, $b = 4.94643$ Å, $c = 5.04817$ Å, $\beta = 90.0000°$} | | | |
| Atom | x | y | z |
| In (2c) | 0.75 | 0.5 | 0.25 |
| Ta (2a) | 0.25 | 0.0 | 0.25 |
| O (8m) | 0.51940 | 0.23869 | 0.94648 |
| \multicolumn{4}{c}{$a = 4.94097$ Å, $b = 5.07150$ Å, $c = 4.90870$ Å, $\beta = 90.38°$} | | | |
| Atom | x | y | z |
| In (2f) | 0.5 | 0.75010 | 0.25 |
| Ta (2e) | 0 | 0.24895 | 0.25 |
| O$_1$ (4g) | 0.23856 | 0.94519 | 0.51851 |
| O$_2$ (4g) | 0.23900 | 0.44598 | 0.47955 |

**Table VI:** Calculated IR modes and pressure coefficients for the low- and high-pressure phases of InTaO$_4$, at ambient pressure and 20.5 GPa, respectively.

| | Low pressure | | | High pressure | |
|---|---|---|---|---|---|
| Mode | ω (cm$^{-1}$) | dω/dP (cm$^{-1}$/GPa) | Mode | ω (cm$^{-1}$) | dω/dP (cm$^{-1}$/GPa) |
| B$_u$ | 148.2 | -0.15 | B$_u$ | 2.83 | 0.60 |
| B$_u$ | 176.8 | 1.41 | B$_u$ | 4.0 | 1.26 |
| A$_u$ | 178.3 | 1.30 | B$_u$ | 40.3 | 3.90 |
| B$_u$ | 178.4 | 0.31 | B$_u$ | 88.9 | 4.70 |
| B$_u$ | 213.6 | -2.24 | A$_u$ | 101.5 | 2.40 |
| B$_u$ | 215.2 | -3.25 | B$_u$ | 178.2 | 0.30 |
| B$_u$ | 258.0 | 1.70 | A$_u$ | 189.5 | 1.80 |
| A$_u$ | 281.7 | -0.25 | A$_u$ | 197.8 | -2.93 |
| A$_u$ | 315.7 | -0.05 | B$_u$ | 228.7 | 2.07 |
| B$_u$ | 332.2 | 3.99 | A$_u$ | 363.5 | 1.27 |
| A$_u$ | 438.1 | 3.88 | B$_u$ | 393.5 | 0.30 |
| B$_u$ | 467.1 | 3.53 | B$_u$ | 451.4 | 2.78 |
| B$_u$ | 493.7 | 5.78 | B$_u$ | 530.2 | 1.60 |
| A$_u$ | 511.0 | 4.65 | A$_u$ | 630.2 | 1.27 |
| A$_u$ | 591.3 | 5.58 | A$_u$ | 640.4 | 3.20 |
| B$_u$ | 615.4 | 5.41 | B$_u$ | 650.1 | 2.40 |
| A$_u$ | 763.3 | 4.10 | A$_u$ | 813.1 | 2.10 |

**Figure captions**

**Figure 1:** (color online) Crystal structure of the low-pressure (top) and high-pressure (bottom) phases of $InTaO_4$.

**Figure 2:** (color online) Selection of XRD measured on $InTaO_4$ at different pressures. Pressures in GPa are indicated on the left-hand side of the figure. (r) denotes a pattern measured upon decompression.

**Figure 3:** (dots) XRD patterns measured at 1.4 GPa (low-pressure phase) and 20.4 GPa (high-pressure phase) for $InTaO_4$. The Rietveld refinements, backgrounds, and residuals are also shown (solid lines). The ticks indicate the calculated positions for Bragg peaks.

**Figure 4:** Pressure dependence for the unit-cell parameters (top) and unit-cell volume (bottom) of the low-pressure phase (solid symbols) and the high-pressure phase (empty symbols). The solid lines are the results of the calculations and dashed lines the EOS fitted from the experimental results. The inset shows the pressure dependence of the monoclinic β angle.

**Figure 5:** Selection of Raman spectra measured at different pressures (indicated in the plot). The asterisks indicate the detection of peaks of the HP phase when this is the minority phase. The dollar symbols indicate the detection of peaks of the low-pressure phase when this the minority phase. Ticks indicate the position of Raman modes for the low- and high-pressure phases. (r) denotes a spectrum measured upon decompression.

**Figure 6:** Pressure dependence of the Raman modes of the low-pressure (circles) and high-pressure (squares) phases. Solid and empty symbols are used alternatively to facilitate the identification of different modes. The solid lines are the results of the linear fits shown in Tables III and IV.

**Figure 7:** (top) Absorption spectra measured at different pressures for the low-pressure phase. The inset shows $E_g$ vs. pressure. The symbols are the experimental results. The solid line represents our calculations and the dashed line is the fit to the experimental results. (bottom) Tauc plot used to determine $E_g$. The dashed line shows the extrapolation of the linear region to the abscissa.

**Figure 8:** (top) Absorption spectra measured at different pressures for the high-pressure phase. The inset compares the pressure dependence of $E_g$ in the low-pressure (squares) and high-pressure (circles) phases. The symbols are the experimental results. The solid lines represent our calculations. (bottom) Tauc plot used to determine $E_g$ in the HP phase. The dashed line shows the extrapolation of the linear region to the abscissa.

**Figure 9:** Total energy vs. volume for the low- and high-pressure phases of $InTaO_4$. The inset shows the enthalpy difference as a function of pressure curves showing the phase transitions reported here. The wolframite phase has been taken as a reference.

**Figure 10:** Band structure of the wolframite phase at ambient pressure (top) and 7.3 GPa (center) and for the monoclinic high-pressure phase at 20.5 GPa (bottom).

**Figure 11:** (color online) Total and partial density of states of the wolframite phase at ambient pressure (top) and 7.3 GPa (center) and for the monoclinic high-pressure phase at 20.5 GPa (bottom).

**Figure 1**

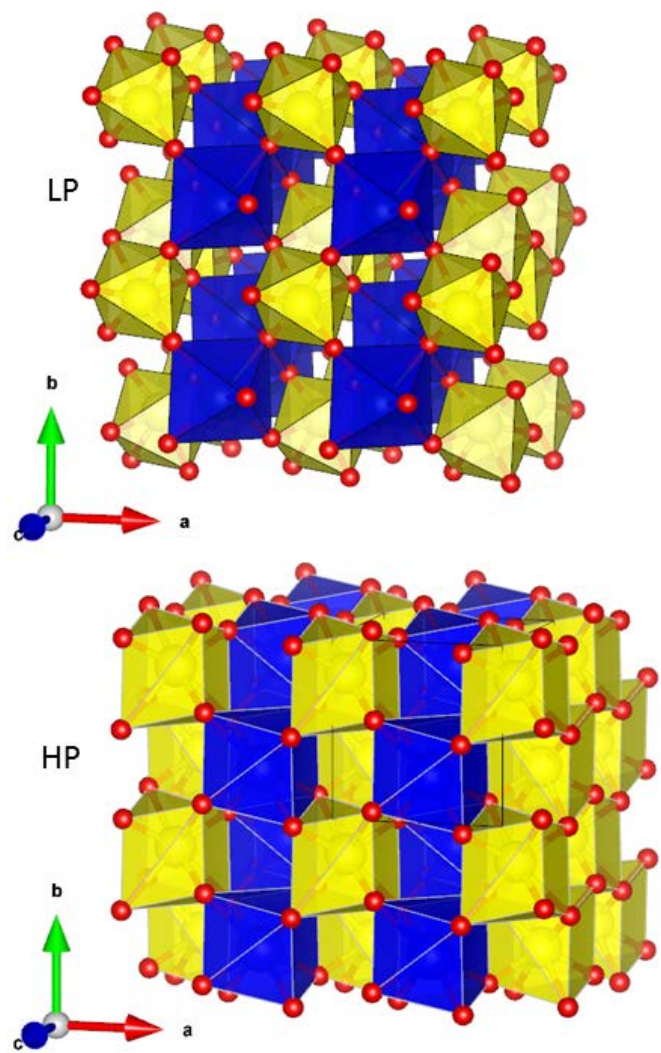

**Figure 2**

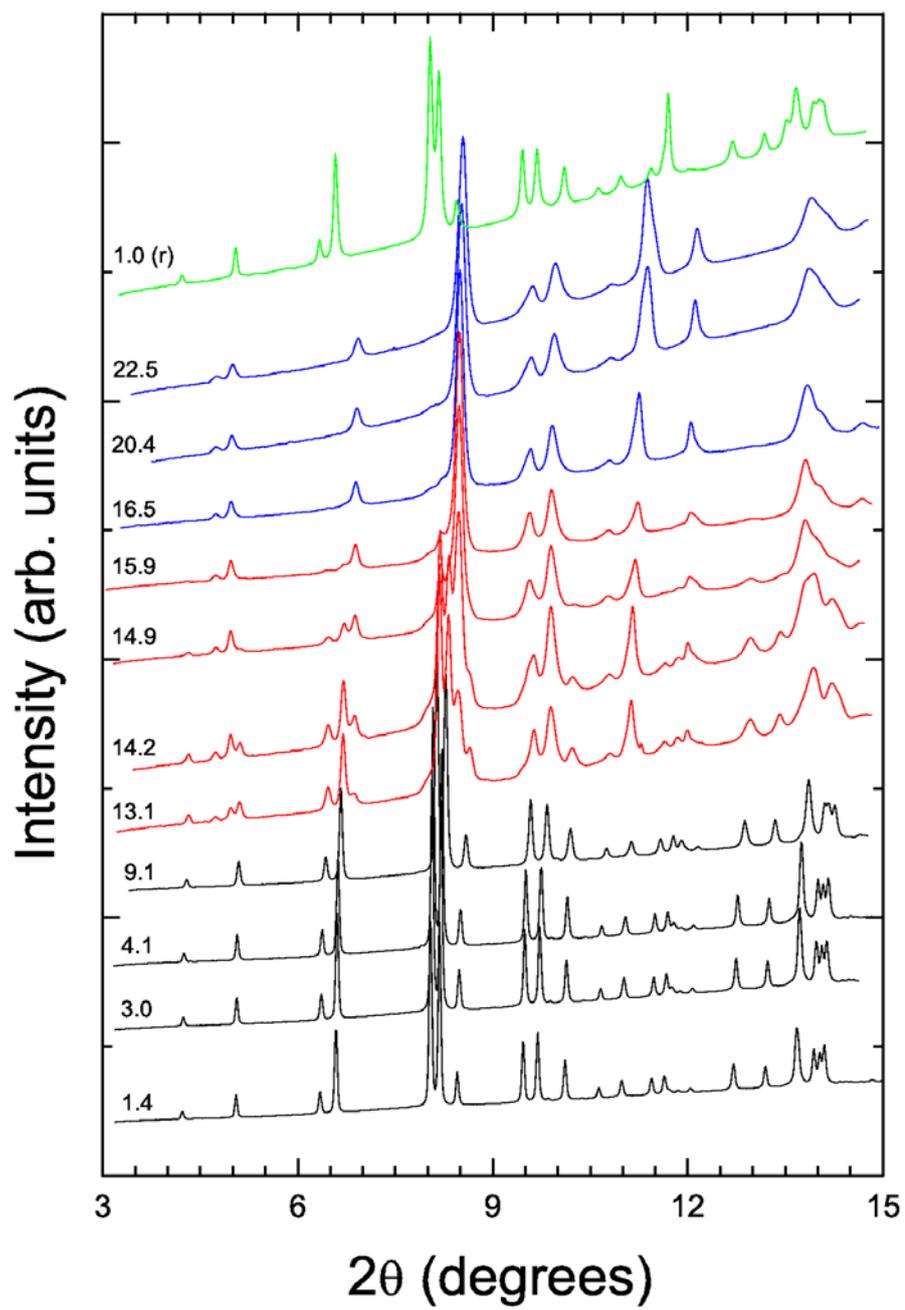

**Figure 3**

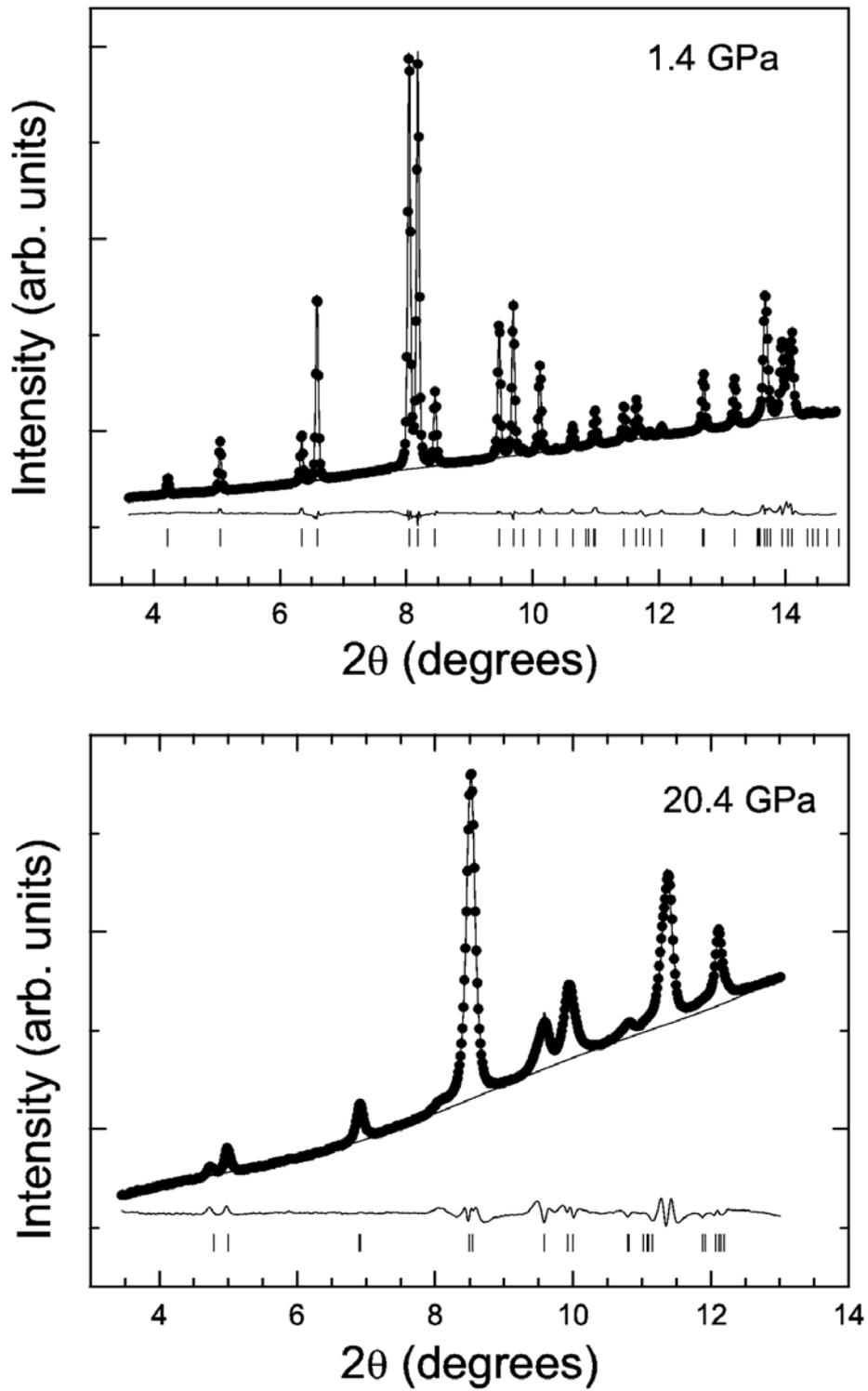

**Figure 4**

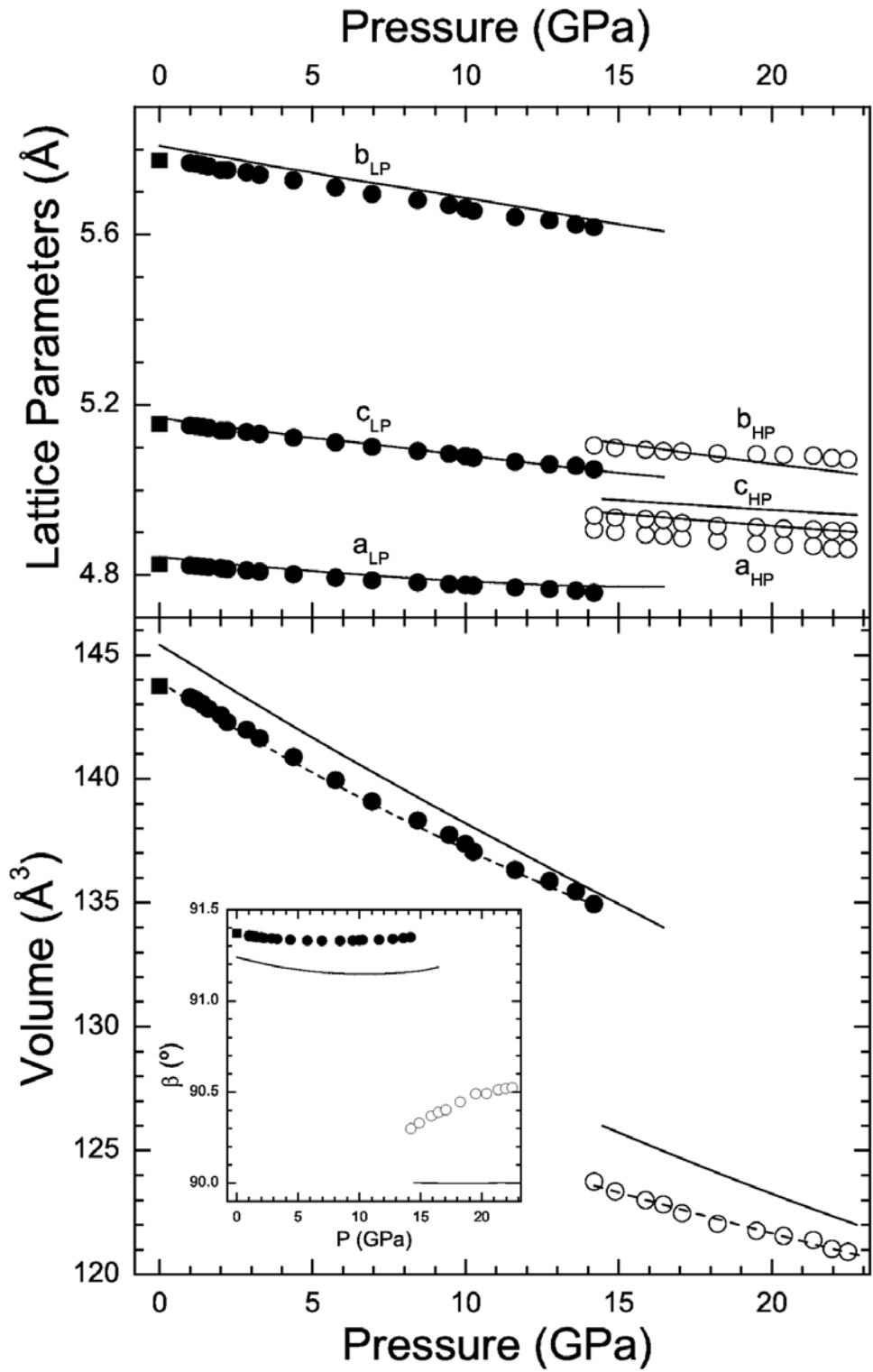

**Figure 5**

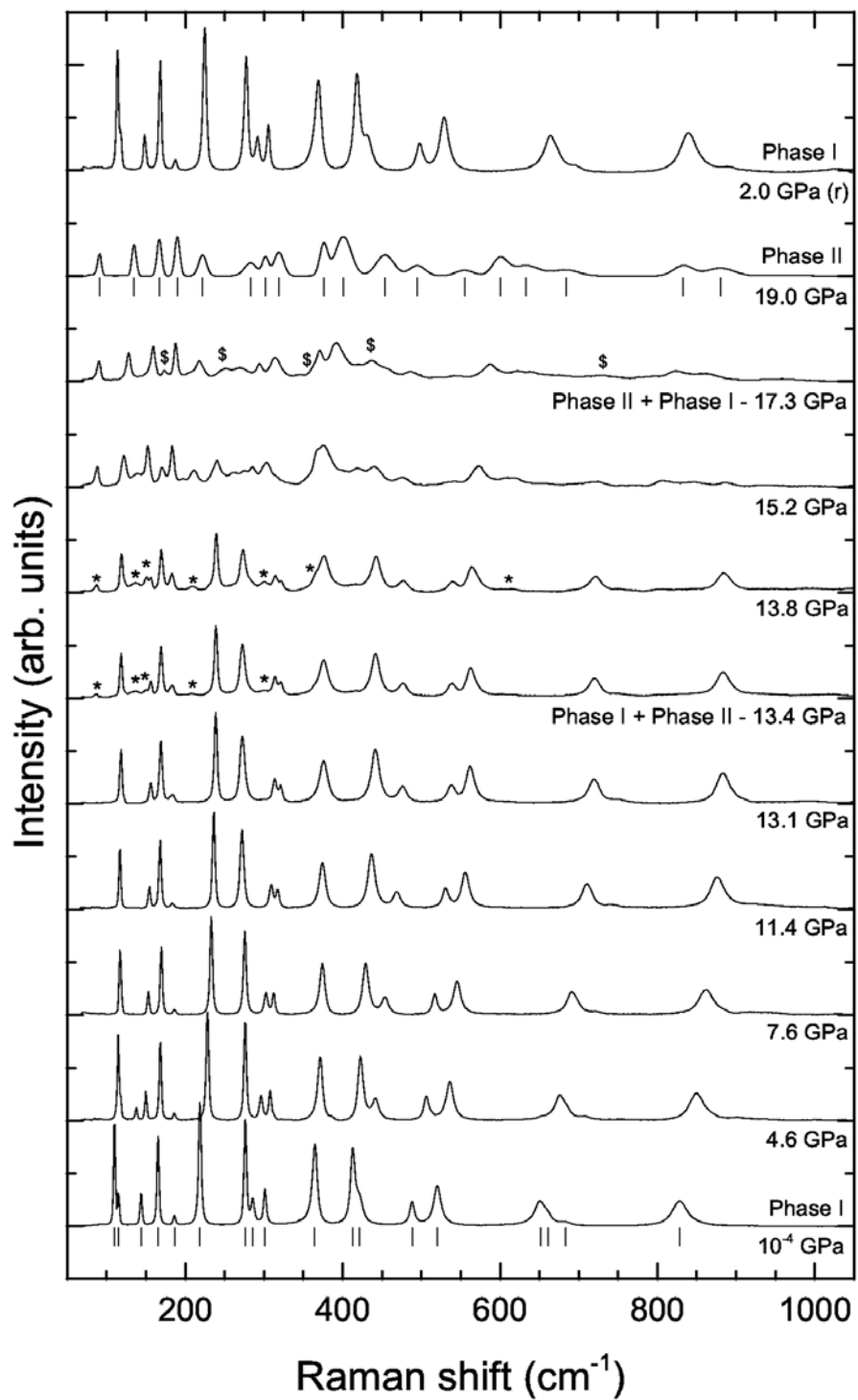

**Figure 6**

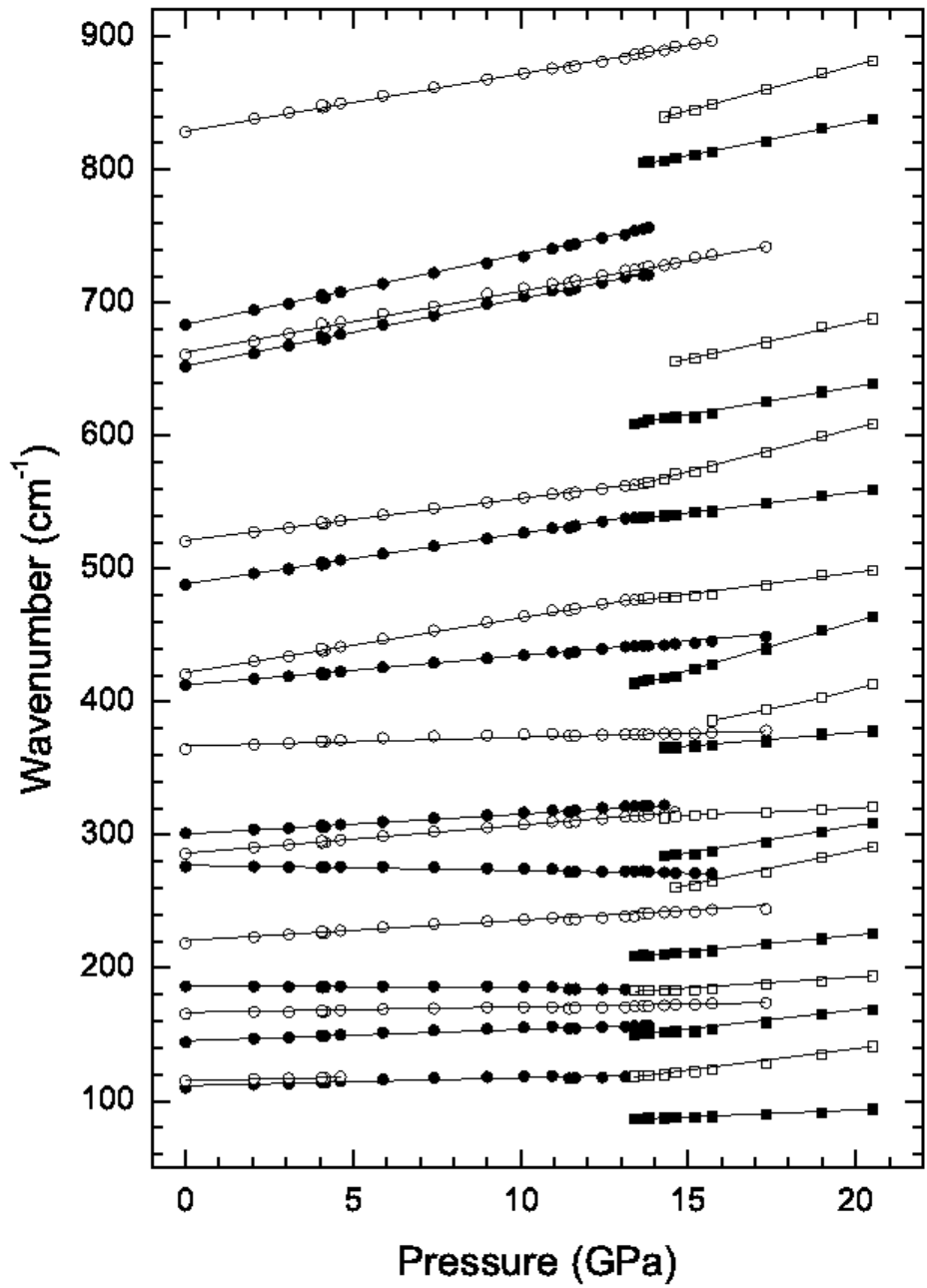

**Figure 7**

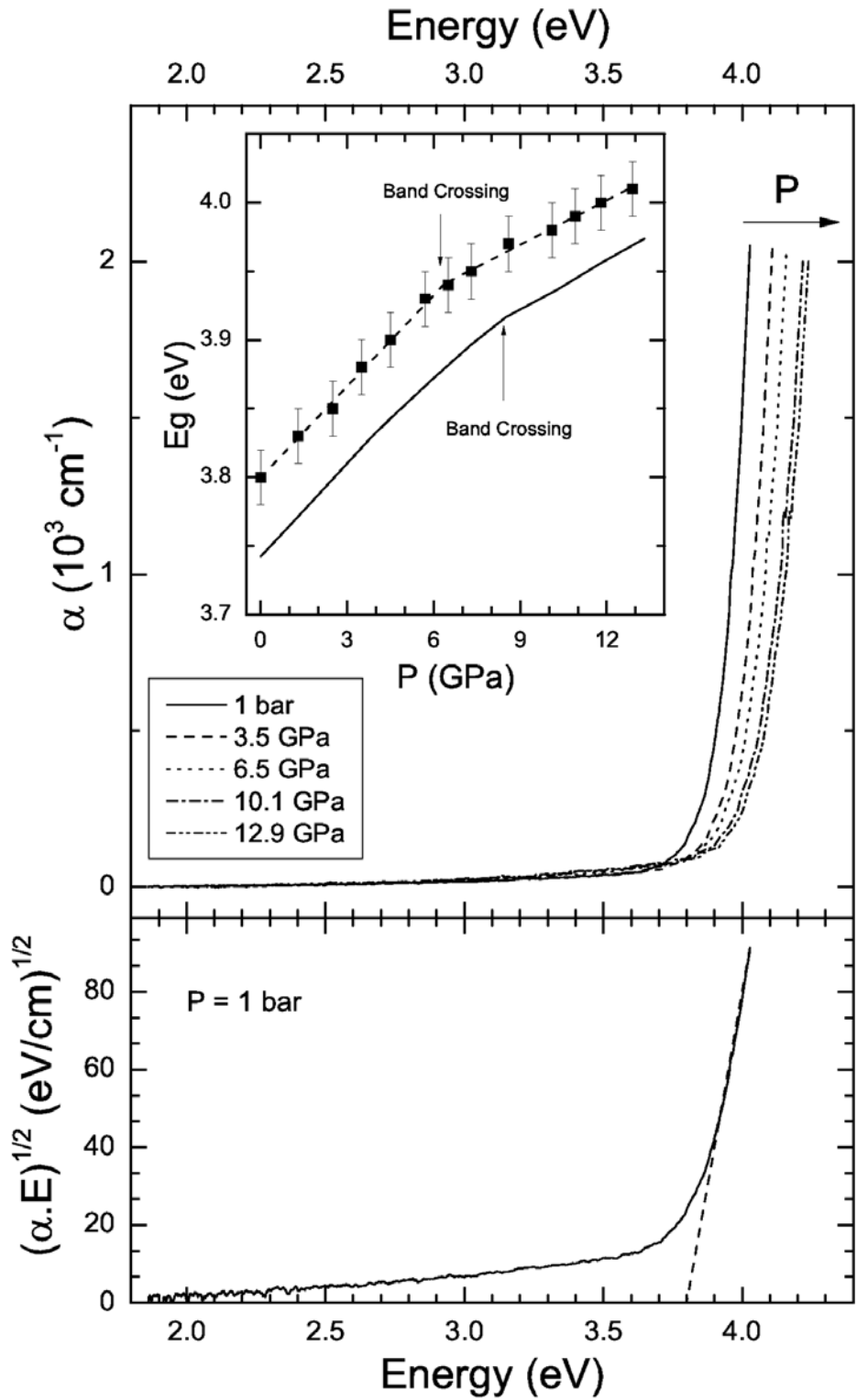

**Figure 8**

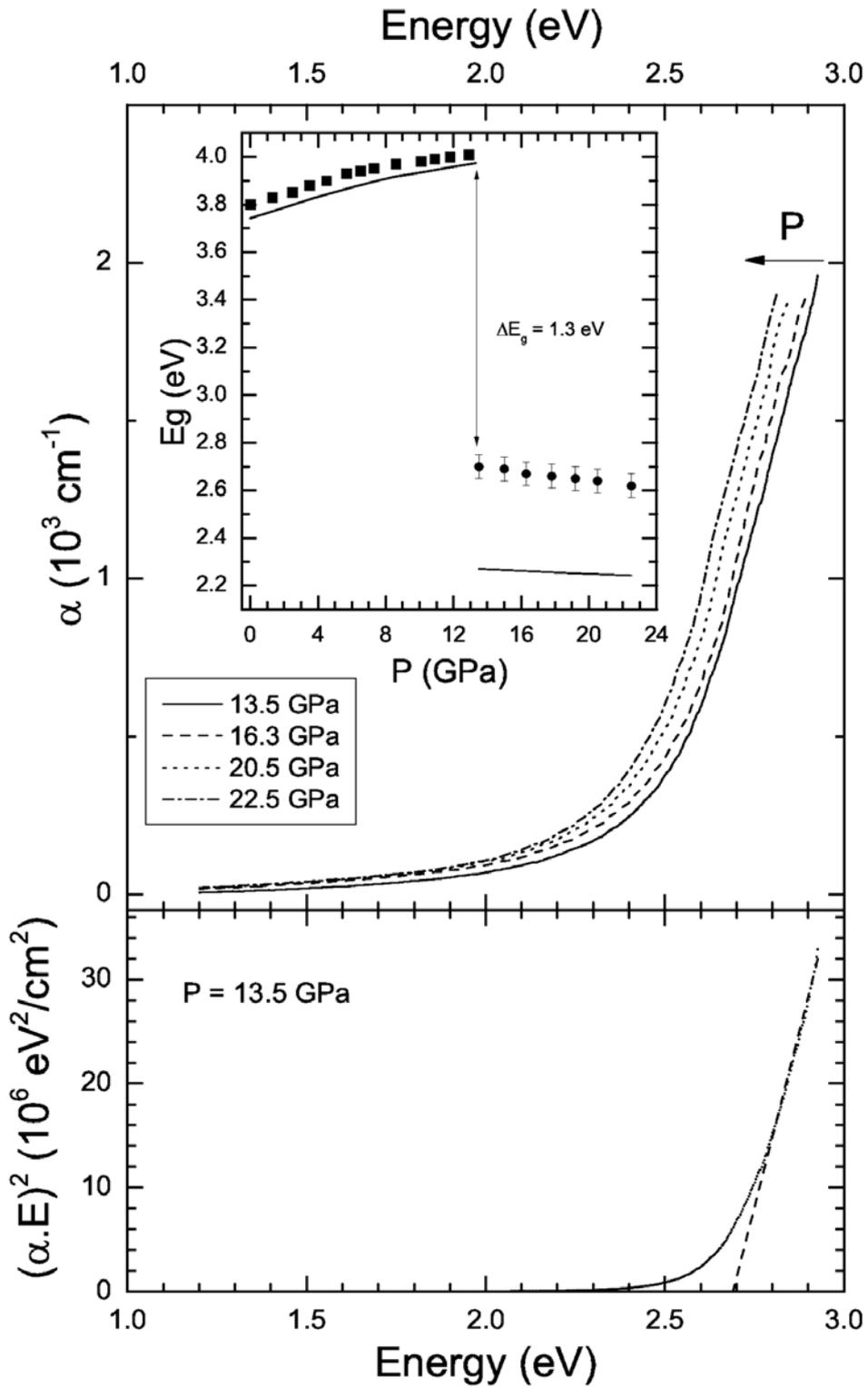

**Figure 9**

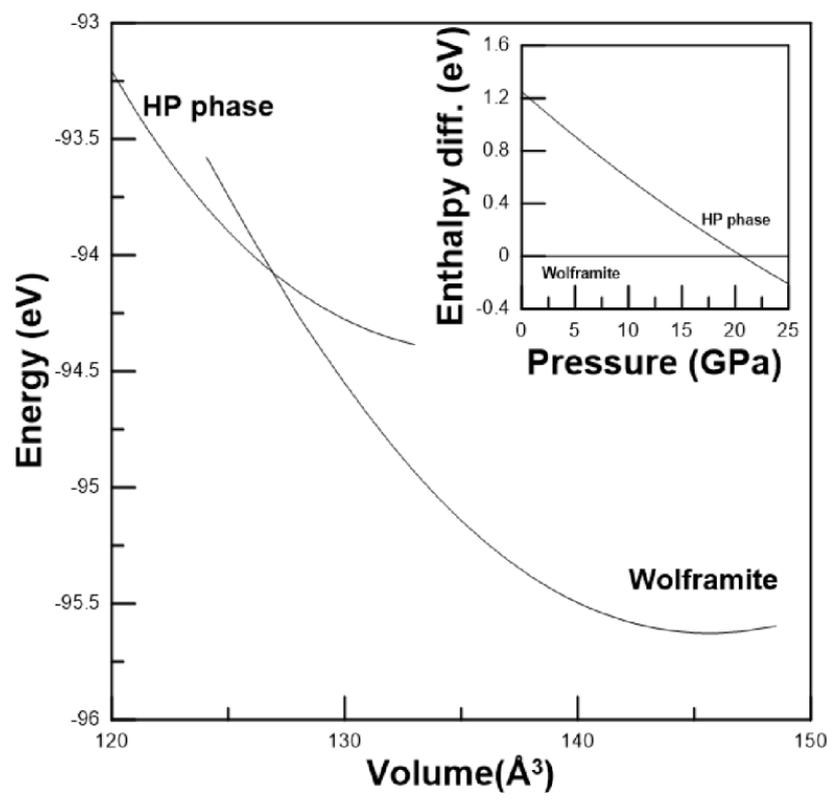

**Figure 10**

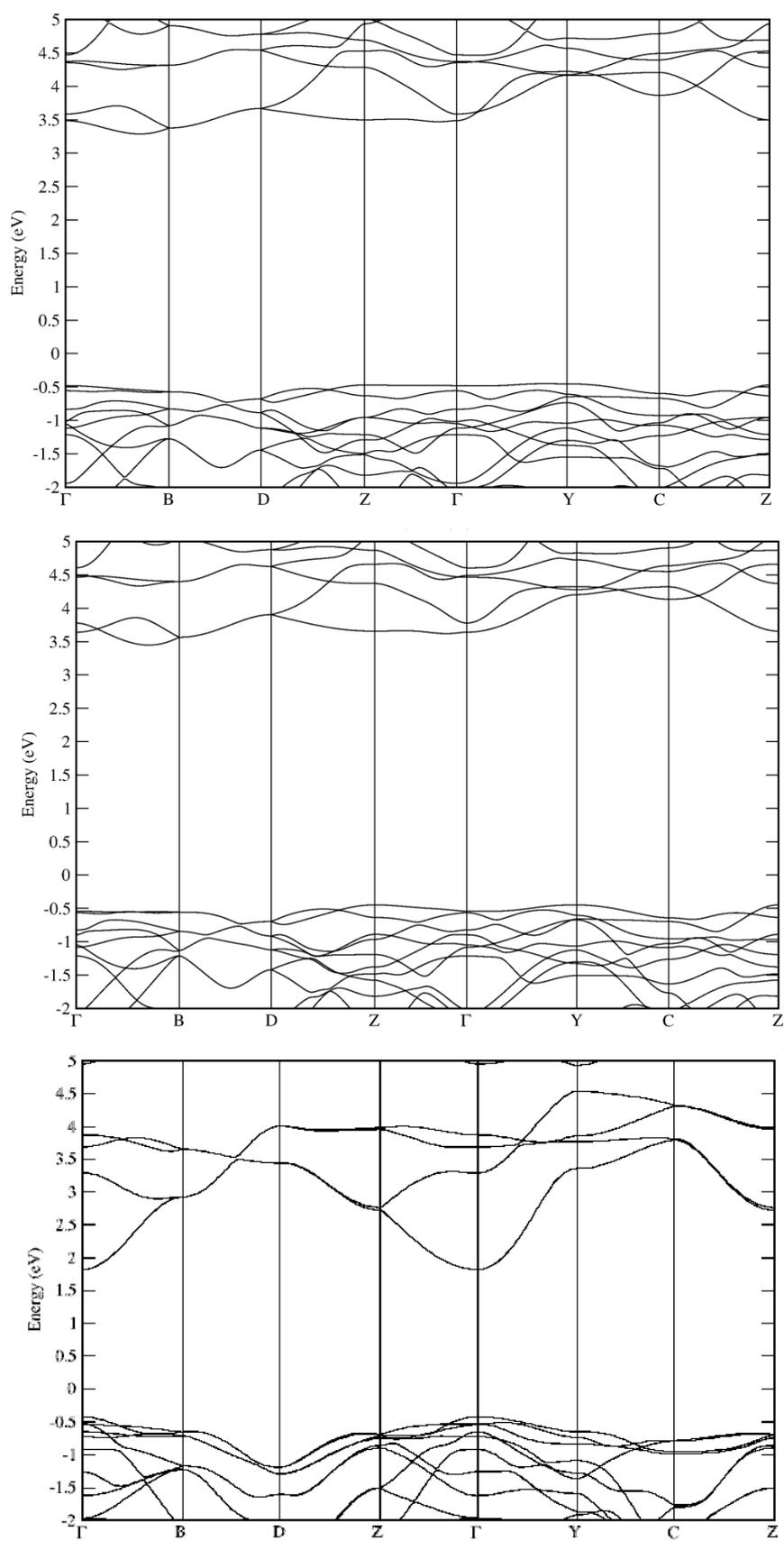

**Figure 11**

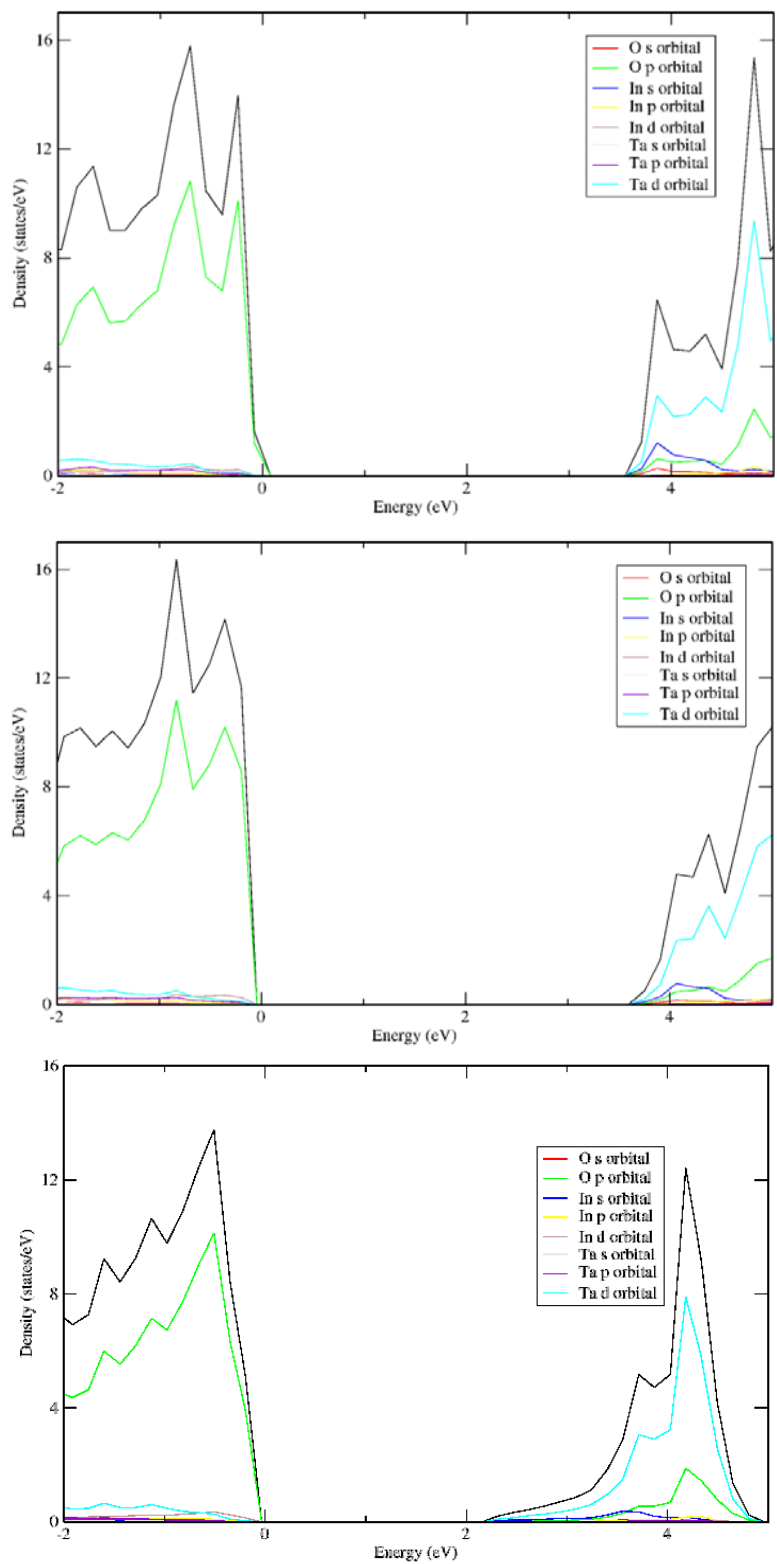